\documentclass[aps, prl, twocolumn, superscriptaddress]{revtex4}
\usepackage{graphicx}
\usepackage{dcolumn}
\usepackage{amssymb}
\usepackage{color} 

\begin{document}

\newcommand{\Obhh}{\Omega_{\mathrm{b}}h^{2}}
\newcommand{\Omhh}{\Omega_{\mathrm{m}}h^{2}}
\newcommand{\As}{A_{\mathrm{s}}}
\newcommand{\ns}{n_{\mathrm{s}}}
\newcommand{\fd}{f_{10}}
\newcommand{\LCDM}{$\Lambda$CDM\ }
\newcommand{\Dchi}{\Delta \chi_{\mathrm{eff}}^{2}}
\newcommand{\ML}{\mathcal{L}_{\mathrm{max}}}

\title{Fitting CMB data with cosmic strings and inflation}

\newcommand{\addressSussex}{Department of Physics \&
Astronomy, University of Sussex, Brighton, BN1 9QH, United Kingdom}

\author{Neil Bevis} 
\affiliation{\addressSussex}
\affiliation{Theoretical Physics, Blackett Laboratory, Imperial College, London, SW7 2BZ, United Kingdom}

\author{Mark Hindmarsh} 
\affiliation{\addressSussex}

\author{Martin Kunz}
\affiliation{D\'epartement de Physique Th\'eorique, Universit\'e de Gen\`eve, 1211
Gen\`eve 4, Switzerland}

\author{Jon Urrestilla}
\affiliation{\addressSussex}
\affiliation{Institute of Cosmology, Department of Physics and Astronomy, Tufts University, Medford, MA 02155, USA}

\date{21 November 2007}

\begin{abstract}
We perform a multiparameter likelihood analysis to compare measurements of the cosmic microwave
background (CMB) power spectra with predictions from models involving cosmic strings. Adding strings to the standard case of a primordial spectrum with power-law tilt $\ns$, we find a 2$\sigma$ detection of strings: $\fd = 0.11 \pm 0.05$, where f10 is the fractional contribution made by strings in the temperature power spectrum (at $\ell = 10$). CMB data give moderate preference to the model $\ns = 1$ with cosmic strings over the standard zero-strings model with variable tilt. When additional non-CMB data are incorporated, the two models become on a par. With variable $\ns$ and these extra data, we find that $\fd < 0.11$, which corresponds to $G\mu < 0.7\times10^{-6}$ (where $\mu$ is the string tension and G is the gravitational constant).
\end{abstract}


\maketitle


\emph{Introduction}.--- 
The inflationary paradigm is successful in providing a match to measurements of the cosmic microwave background (CMB) radiation, and it appears that any successful theory of high energy physics must be able to incorporate inflation. While ad hoc single-field inflation can provide a match to the data, more theoretically motivated models commonly predict the existence of cosmic strings \cite{CS}. These strings are prevalent in supersymmetric D- and F-term hybrid inflation models (see eg. \cite{Lyth:1998xn}) and occur frequently in grand-unified theories (GUTs) \cite{Jeannerot:2003qv}. String theory can also yield strings of cosmic extent \cite{stringtheory}. Hence the observational consequences of cosmic strings are important, including their sourcing of additional anisotropies in the CMB radiation. 

In this letter we present a multi-parameter fit to CMB data for models incorporating cosmic strings. It is the first such analysis to use simulations of a fully dynamical network of local cosmic strings, and the first to incorporate their microphysics with a field theory \cite{Bevis:2006mj, Bevis:2007qz}. It yields conclusions which differ in significant detail from previous analyses based upon simplified models: we find that the CMB data \cite{CMBDATA} moderately favor a 10\% contribution from strings to the temperature power spectrum measured at mulitpole $\ell=10$ with a correponding spectral index of primordial scalar perturbations $\ns \simeq 1$. There are also important implications for models of inflation with blue power spectra ($\ns>1$). These are disfavoured by CMB data under the concordance model (power-law \LCDM which gives $\ns=0.951^{+0.015}_{-0.019}$ \cite{Spergel:2006hy}) and previous work seemed to show that this remains largely the case even if cosmic strings are allowed ($\ns=0.964\pm 0.019$ \cite{Battye:2006pk}). However with our more complete CMB calculations, we find that the CMB puts no pressure on such models if they produce cosmic strings. Our conclusions are slightly modified when additional non-CMB data are included, with the preference for strings then reduced.

\emph{CMB calculations}.--- 
In the combined inflation plus strings case, inflation creates primordial perturbations which evolve passively until today but, in the intervening time period, cosmic strings actively source additional perturbations. Given the small size of the observed CMB anisotropies, the perturbations may be treated linearly and any coupling between those seeded by the two mechanisms can be ignored. The string and inflation perturbations can therefore be evolved via separate calculations, yielding two contributions to the CMB power spectrum that are statistically independent and so are simply added together to give the total power spectrum.

Calculating the cosmic string component presents a challenge because their evolution is non-linear and the string width is very much smaller than their separation at times of importance for CMB calculations. Previous comparisons of the string CMB power spectrum against data have relied upon models which neglect the width, representing local strings as 1D objects and then either evolving them according to the Nambu-Goto equations appropriate for a relativistic string \cite{Contaldi:1998mx} or employing an unconnected segment model (USM) \cite{Wyman:2005tu,Battye:2006pk}. These USMs involve ensembles of unconnected string segments with stochastic velocities and with segments removed to mimic the time dependence of the string density seen in simulations. A third approach is to simulate instead global strings, which do not localize their energy into the string cores. The cores may be left unresolved and field-based CMB calculations \cite{Pen:1997ae} have been used elsewhere \cite{Bouchet:2000hd,Fraisse:2006xc}.

In \cite{Bevis:2006mj, Bevis:2007qz} we used a field-based approach for local strings, via the Abelian Higgs model. We were able to resolve the cores and to reach a string separation of $\sim\!100$ times their width, which we carefully checked to be sufficient to reach a \emph{scaling} regime. This regime, in which the statistical properties of the network scale with the horizon size, is of critical importance as it enables the statistical results to be applied to the later times required in CMB calculations. A great advantage of the field theory is that it naturally includes the decay of the string network into Higgs and gauge radiation, and the resulting backreaction on the network. Thus our CMB calculations for strings are the first to include a consistent mechanism for decay and backreaction.

A feature of field theory simulations is a very low density of string loops \cite{Vincent:1997cx}, in sharp distinction to Nambu-Goto simulations on which the conventional cosmic string scenario is based. Further work is needed to understand the origin of the difference, on which bounds from cosmic rays \cite{Vincent:1997cx} or gravitational wave production \cite{Battye:2006pk} sensitively depend, but CMB calculations depend on the large-scale properties, about which there is broad agreement. Indeed the USM has enough flexibility to approximate our power spectrum: the left hand graph of Fig.\ 1 of the erratum to \cite{Wyman:2005tu} is similar to Fig.\ 13 of \cite{Bevis:2006mj}. However, the USM does not reproduce the detailed shape of the power spectra, nor can it give limits on the string tension $\mu$ without reference to simulations such as ours. Our calculations represent a significant step forward in reliability and accuracy, deserving careful comparison to the data.

\begin{figure}
\vspace{-2.55mm}
\resizebox{\columnwidth}{!}{\includegraphics{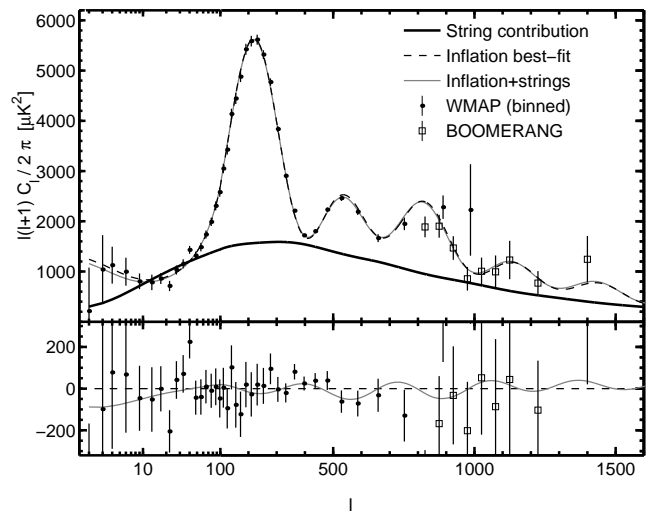}}
\caption{\label{fig:TT} The temperature power spectrum contribution from cosmic strings, normalized to match the WMAP data at \mbox{$\ell=10$}, as well as the best-fit cases from inflation only (model PL) and inflation plus strings (PL+S). These are compared to the WMAP and BOOMERANG data. The lower plot is a repeat but with the best-fit inflation case subtracted, highlighting the deviations between the predictions and the data. Note that the string contribution is identical to that shown in Fig. 14 of \cite{Bevis:2006mj}, but here has a linear horizontal axis for $\ell>100$.}
\end{figure}

\emph{Data fitting approach}.--- 
The form of the cosmic string contribution to the temperature power spectrum is shown in Fig. \ref{fig:TT}, where it is compared to observational data and the best-fit standard inflation model. The normalization of the inflation and string power spectra components are free parameters, with that for strings being proportional to $(G\mu)^{2}$ (where $G$ is the gravitational constant and $\mu$ is the string tension). For Fig. \ref{fig:TT} the normalization of the string component has been set to match the data at multipole $\ell=10$, corresponding to $G\mu=(2.04\pm0.13) \times 10^{-6}$, a factor of 2-3 higher than the corresponding value from previous work \cite{Landriau:2002fx, Contaldi:1998mx, Wyman:2005tu}. Clearly a string component this large is ruled out and we hence introduce the parameter $\fd$, the fractional contribution from cosmic strings to the temperature power spectrum at $\ell=10$.

Recalculating the inflationary component at a particular cosmology takes only a few seconds, but for the string contribution this takes many hours and it therefore appears that a full Markov chain Monte Carlo (MCMC) multi-parameter fit is unfeasible. However, following \cite{Bevis:2004wk}, we fix the form of the string component and vary only its normalization, via $G\mu$. Given that any changes in the cosmological parameters are small and that the strings are sub-dominant, this amounts to a small error in the total inflation plus strings prediction, below the uncertainties in the CMB data \footnote{The \emph{absolute} uncertainties in the string power spectrum calculations are also smaller than those of the data.} and the MCMC results are unaffected. We hence use a version of the standard \mbox{\textsc{CosmoMC}} \cite{Lewis:2002ah} code, modified to incorporate the fixed-form cosmic string component.

We primarily consider four different models: two parameterizations of the primordial power spectrum, both with and without strings. We always allow for variations in the Hubble parameter $h$, the physical baryon and total matter densities $\Obhh$ and $\Omhh$, as well as the optical depth to last scattering $\tau$. We then either take Harrison-Zeldovich (scale-invariant) adiabatic primordial perturbations with amplitude $\As$ or add the additional freedom of a power-law tilt $\ns$: $\As^{2} \rightarrow \As^{2}(k/k_0)^{\ns}$. This yields the two zero-string models which we label as HZ and PL respectively, with PL being the established inflationary concordance model and HZ being a restriction of this: $\ns=1$. We add strings to these two models yielding models HZ+S and PL+S, which therefore have the extra parameter $(G\mu)^{2}$. Then, in the later stages of our discussion, we also consider primordial tensor perturbations and a finite running of the scalar spectral index $d\ns /d\ln k$, but we will assume negligible neutrino mass and flat space throughout.

\begin{figure}
\resizebox{0.55\columnwidth}{!}{\includegraphics{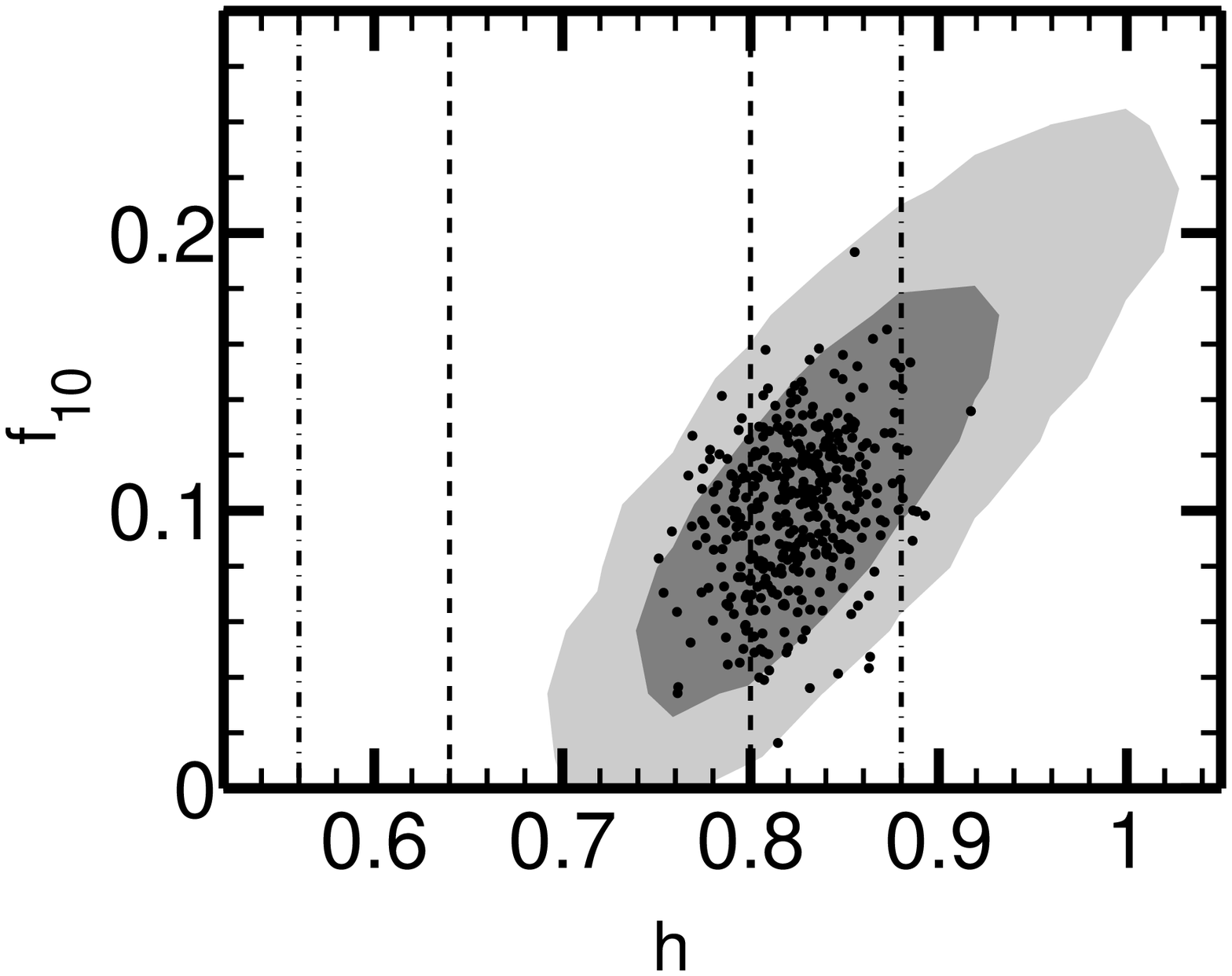}}\hspace{-1mm}
\resizebox{0.449\columnwidth}{!}{\includegraphics{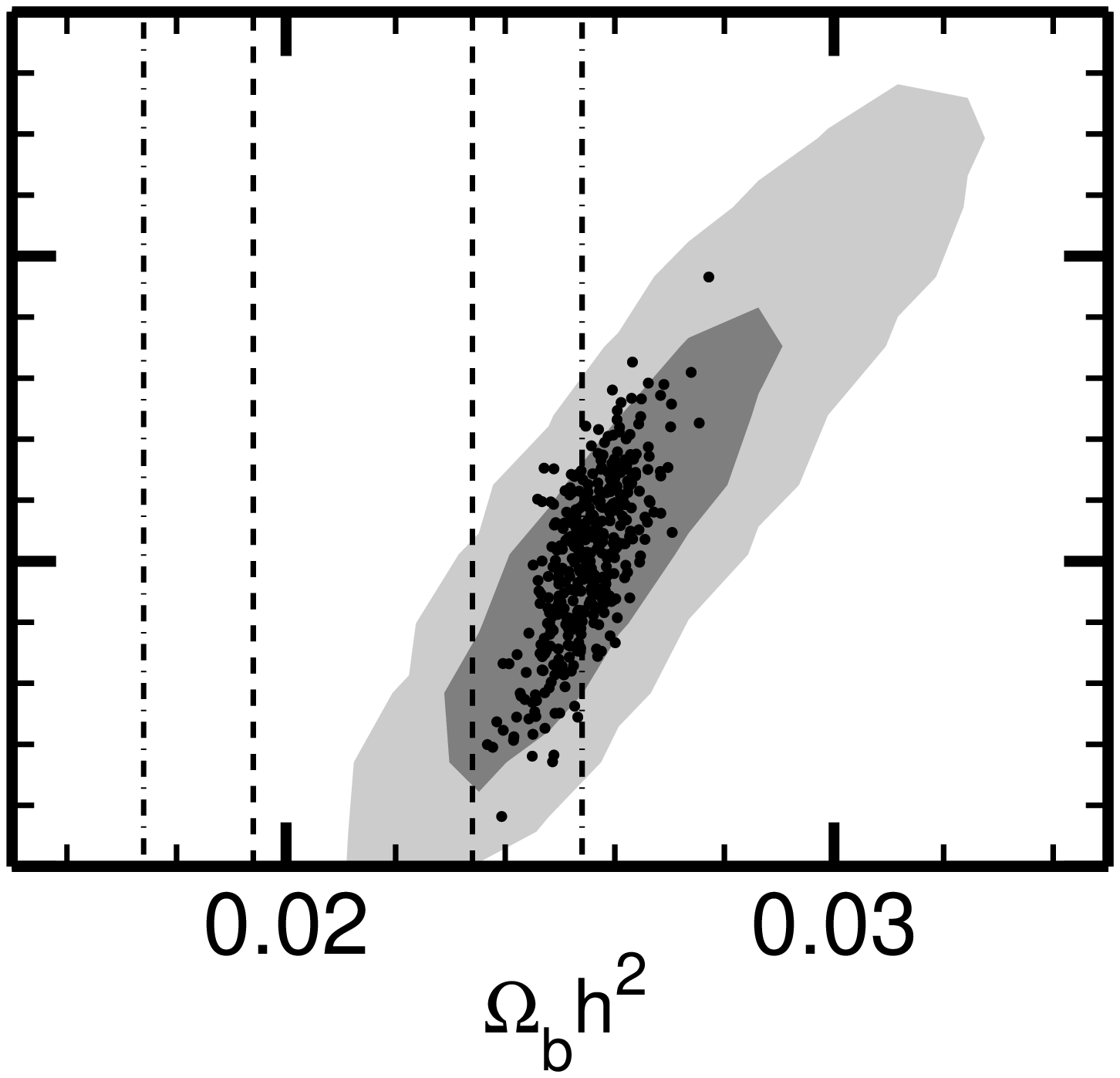}}
\hspace{-1mm}\\
\resizebox{0.55\columnwidth}{!}{\includegraphics{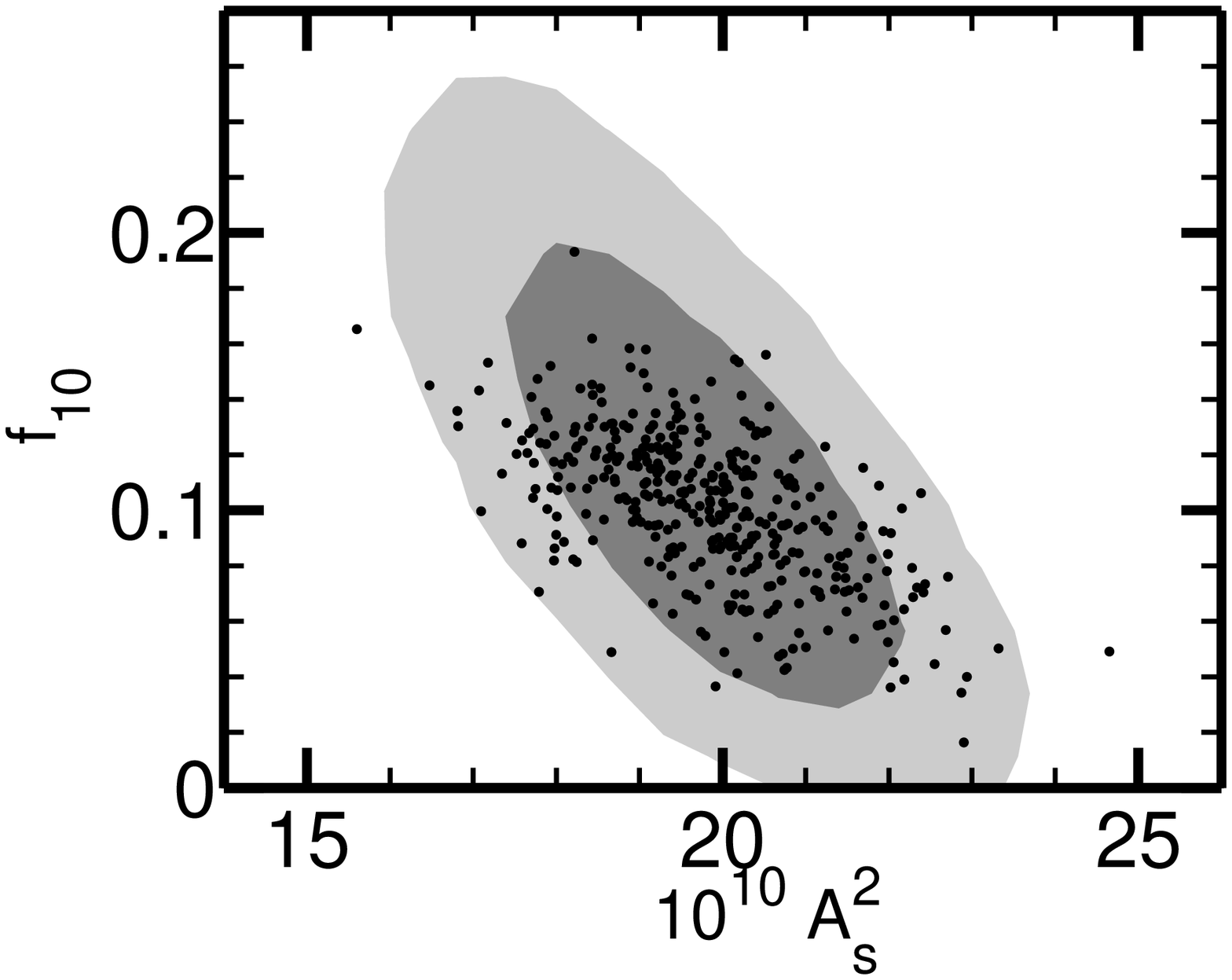}}\hspace{-1mm}
\resizebox{0.449\columnwidth}{!}{\includegraphics{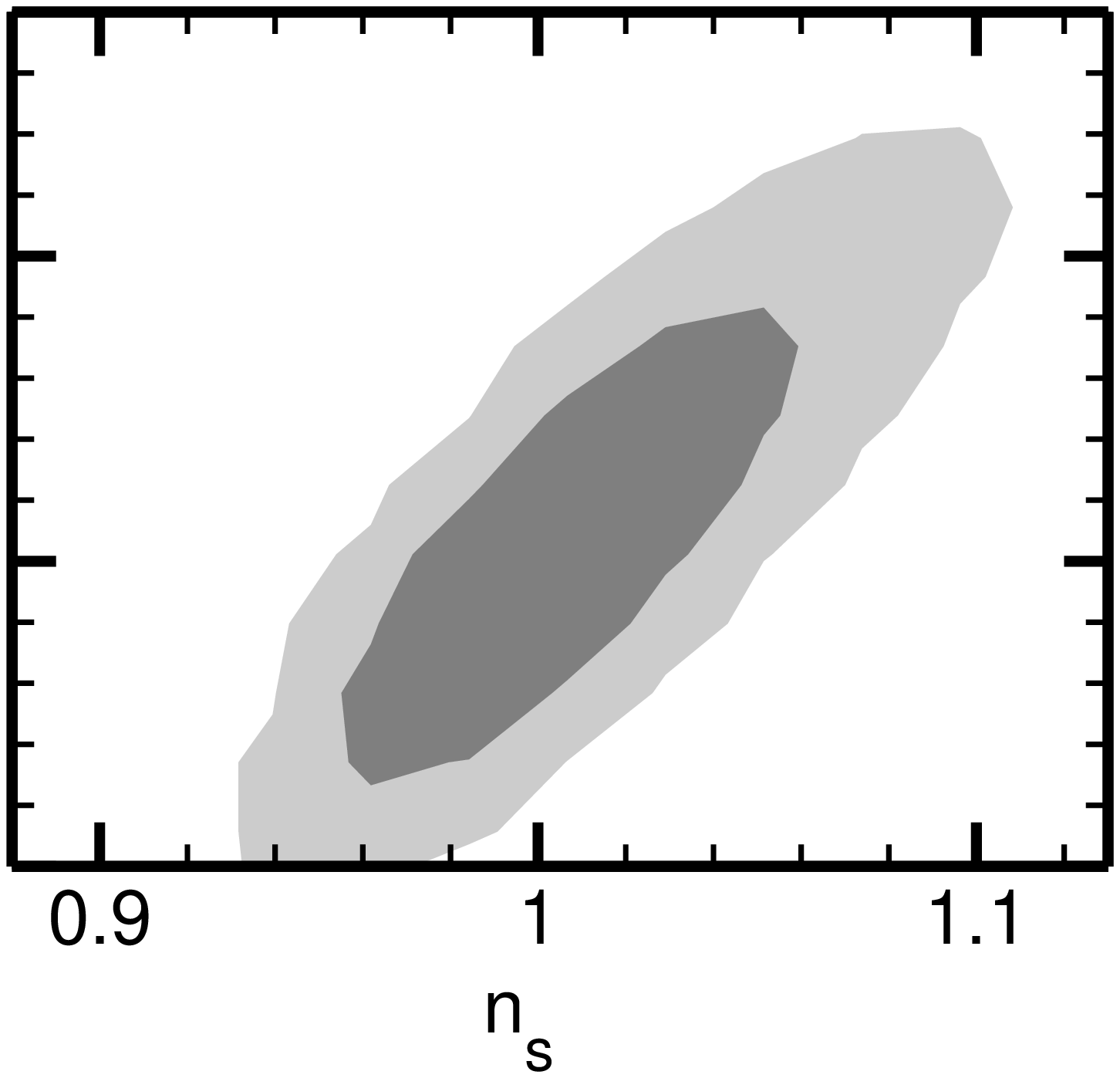}}
\caption{\label{fig:like2D}The 2D marginalized likelihood distributions
from CMB data (only) for $\fd$ versus $h$, $\Obhh$, $\As^{2}$ and $\ns$.
Contours show the $68$ and $95\%$ confidence regions for model PL+S
while the 400 MCMC points indicate the prefered region for HZ+S. The
vertical lines on the $h$ and $\Obhh$ plots show the $68$ and $95\%$
confidence limits from the HKP and BBN measurements.}
\end{figure}

\emph{Results using only CMB data}.--- 
The results when using measurements from the WMAP, ACBAR, BOOMERANG, CBI and VSA projects \cite{CMBDATA} are illustrated in Fig. \ref{fig:like2D}. This shows the marginalized 2D likelihood surfaces for $\fd$ versus $h$, $\Obhh$, $\As^{2}$ and $\ns$ for both HZ+S (points) and PL+S (contours). For PL+S, there is a significant degeneracy, involving primarily these five parameters, that allows large values of $\fd$ to fit the data \footnote{Recalculating the string component at the parameters required for high and low $\fd$ yields an insignificant change, confirming the validity of the method employed.}. The result is $\fd=0.11\pm0.05$, which is a $2\sigma$ detection of strings. It also yields $\ns=1.01\pm0.04$ which is significantly larger than in model PL, or the result of $\ns=0.964\pm0.019$ found in \cite{Battye:2006pk} for PL+S using the USM. 

Figure \ref{fig:TT}(lower) shows the deviations between the best-fit PL+S case, the best-fit PL case and the CMB data. Given that the best-fit PL+S case is given by $\fd=0.099$ and $\ns=1.00$ (see endnote \footnote{The remaining best-fit values are $h=0.82$, $\Obhh=0.0255$, $\Omhh=0.123$, $\tau=0.11$, and \mbox{$\As^{2}=20\times10^{-10}$}.} for the other parameter values), it is clear that not only is $\ns=1$ under no pressure if cosmic strings are included, but it is able to fit the data moderately better than the $\ns=0.952$ best-fit under model PL. Indeed, when the maximum likelihood values $\ML$ are compared via \mbox{$\Dchi = -2\ln(\ML^{\mathrm{PL+S}}/\ML^{\mathrm{PL}})$}, we obtain $\Dchi = -3.9$ at the expense of a single extra parameter. However, as the PL+S best-fit value of $\ns$ is extremely close to one, HZ+S has an almost identical $\ML$ value. Therefore model HZ+S gives $\Dchi = -3.9$ relative to the concordance model with zero cost in terms of the number of parameters.

\renewcommand\arraystretch{1.25}
\begin{table}
\begin{tabular}{|c|c||c|c||c|c|}
\hline
model& no.      & \multicolumn{2}{c||}{CMB only} & \multicolumn{2}{c|}{CMB+HKP+BBN} \\
\cline{3-6}
ID   & param. & $\ \ \Dchi$\ \ & \ \ \ evidence\ \ \     &\ \ $\Dchi$\ \  & \ \ \ evidence\ \ \ \\ 
\hline
HZ   &    5   & $+7.7\;$\ \  & $0.35\pm0.03$     & $+10$\ \ \ \ \ & $0.120\pm0.009$ \\
PL   &    6   &   $\;0$      & $1$                &  0          & 1\\ 
HZ+S &    6   & $-3.9\;\,$   & $7.3\pm1.2$       & $+0.9$      & $0.68\pm0.12$\\
PL+S &    7   & $-3.9\;\,$   & $1.2\pm0.1$       & $-1.6$      & $0.19\pm0.01$\\
\hline
\end{tabular}
\caption{\label{table}The $\Dchi$ and relative Bayesian evidence values for the examined models using the CMB, HKP and BBN data.}
\end{table}

A more complete analysis of the freedom in a model is provided by its Bayesian evidence value \cite{Liddle:2006tc} and Liddle et al. \cite{Liddle:2006tc} have previously used this statistic to demonstrate that WMAP data does not actually rule out model HZ, despite the $\ns=0.951^{+0.015}_{-0.019}$ result returned under the standard model PL. Here, we calculate evidence ratios for our four models using the Savage-Dickey method \cite{SDmethod} with flat priors of \mbox{$0<\fd<1$} and \mbox{$0.75<\ns<1.25$}, giving the results shown in the table. We find that the relative evidence of PL+S to PL is barely distinguishable from unity, as expected for merely a $2\sigma$ detection of strings. However, model HZ+S has a Bayes factor of $7.3\pm1.2$ relative to PL and is therefore moderately preferred. That is, a finite string component is favored by CMB data over a tilted power spectrum and the result: $\fd=0.10\pm0.03$ from model HZ+S is therefore of interest.

\emph{Use of non-CMB data}.--- 
We must also check that these conclusions remain valid when non-CMB data are included and we hence consider that the Hubble Key Project (HKP) yielded $h=0.72\pm0.08$ \cite{Freedman:2000cf}. Further, measurements of deuterium abundance in high redshift gas clouds, combined with big bang nucleosynthesis (BBN) calculations, gives $\Obhh=0.0214\pm0.0020$ \cite{Kirkman:2003uv} and while similar determinations using other light isotopes do not yield global concordance, it is still interesting to consider this measurement also. Figure \ref{fig:like2D} shows these two measurements via vertical lines in the relevant plots and it is clear that they each disfavor large values of $\fd$ in model PL+S. It is also evident that they lower the preference for model HZ+S since the majority of the plotted MCMC plots lie at least $1\sigma$ from these two results.

With these data included, model PL+S now yields $\fd=0.05^{+0.03}_{-0.04}$ or $\fd<0.11$ (95\% confidence) and the $2\sigma$ detection is removed. However, the result of $\ns=0.97\pm0.02$ still does not rule out $\ns>1$ with any confidence (cf. $\ns=0.953\pm0.015$ obtained using the USM with these data \cite{Battye:2006pk}). The Bayes factor for model HZ+S relative to PL is reduced, but only to $0.68\pm0.12$, leaving HZ+S on par with the standard model.

We also incorporate galaxy survey data via the matter power spectrum, although there are uncertainties over the use of such data when strings (or other defects) are included \cite{Bevis:2004wk}. However the CMB constraints, together with our calculated string contribution to the matter spectrum, imply that strings make a negligible contribution to the matter power spectrum on large scales (as is also the case using USM calculations \cite{Wyman:2005tu}). These are the same scales where the zero-string case needs no corrections for non-linearity, which have been questioned in \cite{LSS}. We therefore conservatively include SDSS Luminous Red Galaxy data \cite{Tegmark:2006az} for only $k/h < 0.08 \;  \mathrm{Mpc}^{-1}$ finding that it leaves our results essentially unchanged: $\fd=0.10\pm0.04$ and $\ns=1.00 \pm 0.03$ for PL+S with an evidence value of $7.7\pm0.7$ for HZ+S. Including also data for $0.08 < k/h < 0.2 \; \mathrm{Mpc}^{-1}$ and non-linear corrections \cite{Cole:2005sx} gives $\fd<0.11$, $\ns=0.97 \pm 0.02$ and evidence $0.50\pm 0.05$ but the use of the non-linear regime makes these results less reliable.

Hence, while we await further updates from the observational community regarding these additional data, even with them included, model HZ+S remains competitive relative to PL.

\emph{Tensors and running}.--- 
When the freedom for a non-zero primordial tensor contribution is incorporated as a generalization of model PL, tensor modes give a negligible (and possibly zero) improvement in the fit to CMB data. However they do raise the allowed $\ns$ to $0.98\pm0.03$ \cite{Spergel:2006hy} (CMB only), which is a greater effect than the USM strings of \cite{Battye:2006pk}. As an addition to PL+S, tensors are more preferred but again they increase the allowed $\ns$ values. For the CMB+HKP+BBN case we find $\ns = 0.99\pm0.02$, hence even the BBN data puts no pressure at all on $\ns > 1$ when both strings and tensors are included.

Adding finite $d\ns /d\ln k$ (running) to model PL does give a marginal improvement to the fit, with CMB data preferring a slight negative running \cite{Spergel:2006hy}. This lowers small and large scales relative to intermediate ones and may hence be thought to have a similar effect as strings. However, adding strings smooths out the acoustic peaks and in fact there is little correlation between $d\ns /d\ln k$ and $\fd$. Hence we find that the above results are barely affected by finite running.

\emph{Conclusion}.---  
By including cosmic strings, we find a 6 parameter model with $\ns=1$ that performs better than, or about as well as, the established concordance model, and that the latest data does not necessarily favor $\ns<1$. We also find that, when incorporating the (debatable) deuterium BBN result, the cosmic string contribution is constrained to $\fd<0.11$ or $G\mu<0.7\times10^{-6}$. Even at this level it is likely that cosmic strings will be soon detectable using the B-mode polarization of the CMB \cite{Bevis:2007qz} and we await future data releases with great excitement.

Finally, we note that our bounds have been derived only for classical Abelian Higgs strings with equal vector and scalar particle masses \cite{Bevis:2006mj}, and that, for example, F-term inflation may be more accurately treated using simulations with different values. Similarly, different CMB predictions for strings may be found with $(p,q)$-string networks from string theory \cite{stringtheory}, or from other models such as semi-local strings \cite{semilocal}. A confirmed string detection would open up the challenge of differentiating the models and hence learning a great deal about inflation and high energy physics.

We would like to thank Rob Crittenden, Richard Battye, Andrew Liddle, and David Parkinson for helpful conversations. We acknowledge financial support from PPARC/STFC (N.B., M.H), the Swiss NSF (M.K.), the US NSF, Marie Curie Intra-European Fellowship MEIF-CT-2005-009628 and FPA2005-04823 (J.U.).


\newcommand{\arxiv}[1]{}

\end{document}